\documentclass[a4paper,10pt]{article}
\usepackage[english]{babel}                             %%. they will
\usepackage{amsmath,amsfonts,verbatim,graphicx,float} 
\usepackage{xcolor,soul}
\usepackage{bm}
\def\JJ{\mathfrak J}
\def\demi{\frac{1}{2}}
\def\R{\mathbb R}

\begin{document}

\begin{center}
{\Large The Conformal-Symmetry--Color-Neutrality Connection in Strong Interaction}
\end{center}

\begin{center}
M.\ Kirchbach$^{1,2}$, T.\ Popov$^{2,3}$, and J.-A. Vallejo$^{4,5}$\\
$^1$ Instituto de F\'isica, Universidad Aut\'onoma de San Luis Potos\'i,\\
 Av. Manuel Nava 6, 78290 San Luis Potos\'i, M\'exico,\\
 e-mail:{mariana@ifisica.uaslp.mx}\\
 $^2$Institute for Nuclear Research and Nuclear Energy,\\
Bulgarian Academy of Sciences,\\
 Bul. Tsarigradsko Chaus\'ee 72, 1787 Sofia, Bulgaria\\
$^3$ American University in Bulgaria, \\
Svoboda Bachvarova Str. 8, 
Blagoevgrad 2700, Bulgaria,\\
 e-mail:{tpopov@aubg.edu} \\
$^{4}$     
Facultad de Ciencias, Universidad Aut\'onoma de
     San Luis Potos\'i,\\
     Av. Parque Chapultepec 1570, Lomas del Pedregal\\
 78210 San Luis Potos\'i,  M\'exico\\
$^5$Departamento de  Matem\'aticas Fundamentales, Facultad de Ciencias,\\
 UNED, EC.
 Senda del Rey s/n, 38040 Madrid, Spain,\\
 e-mail:{jvallejo@mat.uned.es}
\end{center}

\begin{flushleft}
{{\bf Abstract:} The color neutrality of hadrons is interpreted as an expression  of  conformal symmetry  of strong interaction, the latter being signaled through the detected  ``walking'' at low transferred momenta, $\lim_{Q^2\to 0}\alpha_s(Q^2)/\pi\to 1 $,  of the strong coupling toward a fixed value ($\alpha_s $ ``freezing''  ).  The fact is that  conformal symmetry admits quarks and gluons to reside on the compactified $AdS_5$ boundary, whose topology is  $S^1\times S^3$, a closed space that  can bear 
exclusively  color-charge neutral  configurations, precisely as required by color confinement. The compactification radius, once employed as a second scale alongside with  $\Lambda_{QCD}$, provides for an  $\alpha_s(Q^2) $ ``freezing'' mechanism in the infrared regime of QCD, thus making  the conformal-symmetry--color-neutrality  connection at low energies evident. In this way, 
perturbative descriptions of processes in the infrared  could acquire 
meaning.  In consequence, it becomes possible to address QCD  by quantum mechanics  in terms of a  conformal wave operator equation, which leads to an efficient description of a wide range of 
data on hadron spectra, electromagnetic form factors,  and  phase transitions.  }
\end{flushleft}

\vspace{2pc}

\section{The puzzle of color confinement }
\label{sec:1}

The non-observability of free quarks,  spin-$1/2$ matter fields bound to hadrons, is puzzling scientists since the very beginning of their  discovery in the 60ies of the past century. Contrary to traditional composite systems, such as molecules, atoms, and nuclei,  
hadrons can not be decomposed into their constituents through interactions with external probes. This peculiarity of strong interactions is related to the existence of three ``strong'' charges carried by the quarks, conditionally termed to as ``colors'', and hypothesized as the fundamental  triplet of the gauge group, $SU(3)_c$,  of strong interaction, a non-Abelian group giving rise to a highly non-linear dynamics among the messengers of strong interaction, the gluons. In one of the cases the dynamics can be such, that the interaction among the quarks grows with the increase of relative distances among them, frustrates their release, and also keeps the net charge neutral (color confinement). Meanwhile, it  systematically weakens with the decrease of the relative distances, thus turning quarks almost non-interacting, though still trapping them in colorless configurations in the interior of hadrons (asymptotic freedom). Over the years, various insights into the mechanisms behind  the non-observability of ``color''  could be gained on the basis of the elaborated  fundamental gauge theory of strong interaction, the Quantum Chromodynamics (QCD), though the first principles provoking
the color confinement remained so far as  open problems. A further  open 
problem  of QCD is that contrary to Quantum Electrodynamics, it is still lacking a properly defined quantum mechanical limit. To progress in that regard,  some fundamental symmetry principles should be emphasized,  and employed in the construction of a quantum mechanical wave equation that describes quark systems interacting by  a potential, whose magnitude would be entirely determined by the fundamental parameters furnishing  QCD. 

\noindent
The symmetry underlying a quantum mechanical interaction problem is always reflected by the quantum numbers of the excited states (the spectrum) of the system under consideration. The spectra of hadrons are dominated, isospin by isospin, by the quantum numbers of the irreducible representations of  $SO(4)$, much alike the levels of an electron bound within the Coulomb potential, though in the hadronic case the level splittings are moderately increasing with the energy, while in the H Atom they  are notably decreasing. This observation, made by several authors \cite{Freedman}-\cite{Afonin3}, hints on a possible relevance of conformal symmetry not only in the electromagnetic but also in the strong interaction sector, at first glance a surprise, given the dependence of the strong coupling, $\alpha_s$   on the (negative) square of the transferred momentum, $(-q^2):=Q^2$. Nonetheless,  the facts that (i) at high $Q^2$  values $\alpha_s (Q^2)\stackrel{Q^2\to \infty}{\longrightarrow } 0$ (asymptotic freedom in the ultraviolet), while (ii) at low $Q^2$ values $\alpha_s(Q^2)/\pi\stackrel{Q^2\to 0}{\longrightarrow} 1$ (conformal window in the infrared) point to the possibility that the dynamics in the two extreme regimes of QCD, the ultraviolet  \cite{Brodski}, and the infrared \cite{Andre}, might be governed  by the conformal symmetry, though expected to be  realized in different fashions.  Specifically in the infrared, the role of the conformal symmetry turns out to be  a pivotal one.  Namely, as it should become clear in due course, it establishes a close link between color-neutrality and the opening of the conformal window. Such occurs upon identifying, in accord with the $AdS_5/CFT_4$ duality, the conformally compactified boundary of $AdS_5$, whose topology is $S^1 \times S^ 3$, as the space hosting the QCD degrees of freedom \cite{JAV_TP_MK_JHEP_2021}. 
This space contains a  closed space-like hyper-surface, on which no single charges can exist, so that systems residing on it are necessarily charge neutral (as required by confinement). As long as in addition  the  isometry of $S^1 \times S^3$ is determined by the  conformal group, $SO(2,4)\supset SO(4)$,    
the potential obtained from  the fundamental solution of the respective 
Lapalcian implements that very symmetry, and can be employed in the design of a conformally symmetric instantaneous potential. 
In this way, the conformal-symmetry (CS)--color-neutrality (CN)  connection in the infrared regime of QCD can be established. 
It is the purpose of the present contribution to briefly review recent progress on that topic. The text is structured as follows. In the next section we briefly outline the genesis of the  CS-CN connection concept in the infrared. In Section 3 we discuss the geometric aspects of color confinement and its relationship to conformal symmetry as introduced via the $AdS_5/CFT_4$ duality. Section 4 is devoted to an alternative interpretation of the CS-CN connection from the perspective of the Jordan algebra $\JJ_2^C$, with the aim of hitting the road towards  generalization to higher dimensions. The text closes by a concise summary section.

\section{Spectroscopic evidence for conformal symmetry of QCD}

Besides the freezing of the strong coupling in the infrared, further evidences for the relevance of the conformal symmetry for hadron physics are independently  provided by data on hadron scatterings  and hadron spectra. In particular, relativistic two-body scattering amplitudes
 are well described in terms of exchange between the particles of physical entities that transform as
 irreducible representations of the Poincar\'e group \cite{Toller1}. Such representations inevitably emerge in the decomposition of the direct products of the representations  describing the incoming and outgoing particle states. 
 This relevance of conformal symmetry for hadrons has been noticed already in the early days of Regge's theory by
 ~\cite{Freedman}{}-\cite{Frazer}, and Regge trajectories with $O(4)$ symmetric poles have been considered \cite{Jones}. 
Furthermore, a dynamical conformal symmetry  approach to the description of hadronic  electromagnetic form factors has been developed for example in \cite{Barut1}.   More  hints come from the hadron spectra, both baryonic and mesonic,  whose quantum numbers are markedly dominated by $SO(4)$ irreducible representations, easily recognizable as finite  sequences of states consisting of  $K$  parity pairs of raising spins, terminating by a parity singlet state of highest spin, with $K$ standing for the value of the four-dimensional angular momentum \cite{Mukh}-\cite{Afonin}. Specifically in \cite{Afonin3} the possible relation of new
 data on light meson spectra reported by the Crystal Barrel Collaboration to conformal symmetry of massless QCD has been first indicated. After the 70ies, the conformal symmetry of hadrons has been addressed in the literature  sporadically and treated by purely algebraic means, while for the practical purposes of continued data evaluation,  potential models based on different Lie symmetries and depending on a large number of free parameters, have been favored. Within this context, it appears important to find a method  to design a potential model which implements the conformal symmetry, depends on same  parameters as QCD,   and which allows for an immediate evaluation of a variety of data. Such a model has been elaborated over the years in the series of articles  \cite{MK_MPLA2000}, \cite{TQC}-\cite{MK_Formfactors}. At first, in \cite{MK_MPLA2000}, a two-parameter empirical mass formula has been suggested which resulted quite adequate for the description of the excitation energies of all light flavor hadrons. This formula reads.  
\begin{equation}
M_{\sigma'}-M_\sigma=m_1\left( \frac{1}{\sigma^2}-\frac{1}{\sigma'\, ^2}\right)  + \frac{1}{2} m_2
\left( \frac{\sigma'\, ^2-1 }{2} -\frac{\sigma^2-1}{2}\right), \quad \sigma=K+1.
\label{mss_frml}
\end{equation}
Here, $\sigma  $ plays a r\'ole similar to the principal quantum number of the H Atom and describes the well known $(K+1)^2$-fold degeneracies of states in a level, viewed as $SO(4)$ irreps. The two parameters $m_1$ and $m_2$ take the values of $m_2=70$ MeV, $m_1=600$ MeV for both nucleon and $\Delta$ excitations. Later this formula has been reported  in \cite{TQC} to  be interpretive in terms of the eigenvalues, $\epsilon_{K\ell}$,  of the trigonometric Rosen-Morse (tRM) potential (here in dimensionless units) and given by,
\begin{equation}
V_{tRM} =\frac{\ell (\ell +1)}{\sin^2 \left(\frac{r}{d}\right)} -2b\cot \left( \frac{r}{d}\right).
\label{tRM}
\end{equation}
Here, $d$ so far has been treated as a matching length parameter to the relative distance $r$, while the $V_{tRM}$  spectrum reads,
\begin{equation}
\epsilon_{K\ell}=-\frac{b^2}{(K+1)^2} +(K+1)^2, \quad \epsilon_{K\ell}\sim M_{\sigma}.
\label{tRM_spctr}
\end{equation} 
Upon resolving the associated Scr\"odinger equation, be it with a linear, or quadratic energy,  a surprising result has been obtained regarding the wave functions, which turned out  to express in terms of some real orthogonal polynomials, which have been  entirely absent from the standard  mathematical physics literature available at that time. Later on, these polynomials have been identified in \cite{Raposo} with the Routh-Romanovski polynomials, scarcely covered by the specialized mathematical literature. Next, the case could be made in \cite{MK_PRD10} that upon a suitable change of variables, the wave equation with $V_{tRM}$ can be transformed to a quantum motion on the three dimensional hyper-sphere, $S^3$. In so doing,  $r$ acquires meaning of the arc length, $r \longrightarrow \stackrel{\frown}{r}$, of a great circle, measured from the North pole of $S^3$,  $d$ becomes the sphere's hyper-radius, $R$, and $(r/d)$ takes the part of an angular variable $\chi$, identical to the second polar angle parametrizing $S^3$ in global geodesic coordinates. Within this context, the $\ell (\ell +1)/\sin^2 (r/d)$ term in (\ref{tRM}) starts playing the part of the centrifugal term on $S^3$, while the cotangent potential solves the Laplace equation on $S^3$, much alike as the Coulomb potential solves the Laplace equation in the 3D flat space. As long as the Laplacian shares the symmetry of the isometry group of the manifold on which it acts, $SO(4)$ in our case, so do the solutions of the Laplace equation, which explains the $SO(2,4)\supset SO(4)$  patterns of the $V_{tRM}$ eigenvalues.  Finally, in \cite{KC2016}  the constraint to  neutrality imposed by  the hyper-spherical manifold  on the total charge of systems placed on it was also addressed. 
In effect, the conformal-symmetry--color-neutrality connection could be revealed. Moreover, via this connection  
the solution of the Laplace equation on $S^3$ could be related to a potential, obtained in \cite{Belitsky}  from Wilson loops with cusps at the North and South poles on $S^3$, which allowed us to express the potential magnitude, $b$, as $2b=\alpha_sN_c$, where $N_c$ stands for the number of colors. In this fashion, the $V_{tRM}$ potential parameters could be directly linked to the fundamental parameters of QCD.
In this parametrization, the potential under discussion has been successfully used in the description of a variety of hadron physics phenomena, ranging from meson spectra \cite{KC2016}, over
 nucleon electromagnetic form factors \cite{MK_Formfactors}, and more recently, to heavy flavored mesons \cite{Ahmed}, and thermodynamic properties of quantum meson gases \cite{AramDavid}.  Especially in the latter case, a quantum gas of charmonium was shown to suffer a phase transition to a Bose-Einstein condensate at Hagedorn's value of the critical temperature.
However, so far the question on  the origin  of the curved hyper-spherical manifold had remained  pending. As a  preliminary hypothesis, it has been seen as the hyper-sphere located at the waist  of a four-dimensional hyperboloid, ${\mathbf H}_1^4$, of one sheet, emerging in the time-like foliation of  space-time in  $dS_4$ 
Special Relativity, whose space-like region has been conjectured in \cite{KC2016} as the internal space of hadrons.
The aforementioned question has found a more assertive answer in \cite{JAV_TP_MK_JHEP_2021},
 presented in the subsequent section.

\section{The geometric foundations of the CS--CN connection}
Within the context of contemporary fundamental concepts, $p$-dimensional space-like  spherical manifolds, $S^p$, appear at the compactified  boundaries of $S^1\times S^p$ anti-de Sitter spaces in $(p+2)$ dimensions \cite{Nicolai}. Especially the $S^3$ sphere of our interest appears at the compactified $AdS_5$  boundary, a space of fundamental interest to QCD from the perspective of the $AdS_5/CFT_4$ gauge-gravity duality. In \cite{JAV_TP_MK_JHEP_2021} it has been demonstrated  in detail that this boundary is topologically equivalent  to the compactified Minkowski space time \cite{PR}. Also there, the interest in the compactified Minkowski space as an internal space of the strong interaction degrees of freedom has been formulated for the first time. This interest is motivated by the observation that on such spaces, charges are forced to appear in pairs of vanishing total charge, much alike the appearance of  color charges in mesons.  The argument goes as follows. Be $M$ a compact manifold with finite volume
$\mathrm{vol}(M)$, which is  equipped by any Riemannian metric $g$. Then a
fundamental solution to the associated Laplacian $\Delta$ is any function $G:M\times M\to \R $ satisfying, for any fixed values of $y\in M$, the equation
\begin{equation}\label{eq28}
\Delta_xG(x,y)=\delta_y-\frac{1}{\mathrm{vol}(M)}\,.
\end{equation}
Alternatively, on  $S^3$ another definition of fundamental solutions to the Laplacian can be given, namely,  taking a base point and its antipodal at the same time, leads  to  \cite{cohl}
\begin{equation}\label{cohl}
\Delta G(x,x')=\delta (x,x')-\delta (-x,x')\,,
\end{equation}
which is nothing but the equation for the well known dipole Green function.
{}For the case of $M$ chosen as  $S^3$, and parametrized by global geodesic hyper-spherical coordinates
$(\chi,\theta,\varphi)$, one has   $\mathrm{vol}(M)=2\pi^2$, and a direct computation
\cite{Brigita} shows that the fundamental solutions at the poles 
$\chi =0$, and $\chi=\pi$, are  given as,
\begin{eqnarray}
G_0(\chi,\theta,\varphi)=\frac{1}{4\pi^2}(\pi-\chi) \cot \chi\,, &&
G_\pi(\rho,\theta,\chi )=-\frac{1}{4\pi^2}\chi  \cot \chi \,,
\label{cotsolfund}
\end{eqnarray}
their dipole combination being,
\begin{equation}
G_\pi (\chi,\theta,\varphi)-G_0(\chi,\theta,\varphi)=-\frac{1}{2\pi}\cot\chi,
\end{equation}
and thus leading to the cotangent potential in (\ref{tRM}). Therefore, the potential in (\ref{tRM}) can be interpreted to be due to the charge neutral configuration of a  $2^1$ pole, 
residing on $S^3$, such as quark-anti-quark. The charge neutrality allows existence on $S^3$ of any $2^n$ poles. Furthermore, in employing the compactification radius, $R$, as another scale along $\Lambda_{QCD}$, and upon reparametrizing  $Q^2c^2$ as  $\left( Q^2c^2 +\hbar^2 c^2/R^2\right)$, this  in the spirit of \cite{Richardson},  removes the logarithmic divergence of the strong coupling at origin according to,  
\begin{eqnarray}
\frac{\alpha_s(0)}{\pi }&=&\lim_{Q^2\to 0} 4 
\left({\beta_0 \ln \left(\frac{      Q^2 c^2}{\Lambda^2_{QCD}} +
    \frac{\hbar^2c^2}{R^2\Lambda_{QCD}^2} \right)}\right)^{-1}\nonumber\\
  &\to& 4\left( {\beta_0 \ln \left(\frac{\hbar^2c^2}{R^2\Lambda_{QCD}^2} \right)}\right)^{-1}. 
\end{eqnarray}
In this fashion, the conformal-symmetry--color-confinement connection at low energies becomes evident, and avenues  open towards perturbative treatments of hadron processes in the infrared. The compactification radius extracted from data on light mesons \cite{KC2016}  is $R=0.58$ fm, while for the charmonium it is $R=0.56$ fm \cite{Ahmed}, i.e. it seems maintains an universal meaning far beyond the infrared.

\section{The Jordan algebra view on conformal symmetry }
So far we have emphasized on the conformal group as isometry of the $S^1\times S^3$ space. Here we focus on an important aspect of its algebra, $so(2,4)$, namely, on its property of being patterned after the 3-graded Lie algebra
\begin{equation}
 \mathfrak g_{+1} \oplus \mathfrak{g}_0 
\oplus \mathfrak {g}_{-1}\simeq so(2,4),
\label{3-gdrd}
\end{equation}
where the Abelian sub-algebras  $\mathfrak g_{-1}$, and $\mathfrak g_{+1}$ are defined in their turn by the  commutators of the components $P_\mu$, of the four-momentum operator, and by the components, $K_\mu$ of the operator of special conformal transformations, respectively.  The generators $P_\mu$ and $K_\mu$               are related to each other via the conformal inversion $I$ as, $K^\mu = I P_\mu I:=P_\mu^\dagger \ , $ with  $I(x^0, \bm x)= \left(\frac{x^0}{x^2}, -\frac{ \bm x}{x^2}\right)$  acting as an involution, $I^2=1$.  Moreover, the commutators of the 
$\mathfrak g_{+1}$ and $\mathfrak g_{-1}$ elements are given by,   $\demi  [K^\mu, P_\nu]={M^\mu}_\nu - \delta^\mu_\nu D  \ $,  and recover the algebra of the Lorentz group, whose generators are $M^\mu\, _\nu$, with an associated to it grading operator,  $D\in \mathfrak g_0 $, which acts as a dilatation, $[D, g ]= k g $,  for any $g \in \mathfrak g_{k}  $, with $k=\pm 1$. This property of the conformal group algebra allows one to link it to Jordan triple systems and pairs of Jordan triples as they appear in the special Jordan algebra, $\JJ_2^C$,  of the $2\times 2$ Hermitian matrices. As a reminder \cite{WBeltram}, a special Jordan algebra is a non-associative algebra of a vector space $\JJ$ over a field, whose multiplication, $\circ$, satisfies the commutative law,  $x\circ y=y\circ x$, and the Jordan identity, $(x^2\circ y)\circ x =x^2 \circ (y\circ x)$. A Jordan triple system (JTS) is a  vector space $\JJ$ endowed with a Jordan triple product, {i.e.,}  a trilinear map $\lbrace \ ,\ ,\ \rbrace: \JJ\times \JJ \times \JJ \longrightarrow \JJ $, 
 satisfying  the symmetry condition,
$\lbrace u,v,w \rbrace =\lbrace w,v,u \rbrace$, together with the identity
\begin{equation}
\lbrace u,v,\lbrace w,x,y  \rbrace\rbrace=
\lbrace w,x,\lbrace u,v,y  \rbrace\rbrace +
\lbrace w, \lbrace u,v,x  \rbrace, y\rbrace -
\lbrace \lbrace v,y,w  \rbrace x,y\rbrace .
\label{5fld}
\end{equation} 
The latter relation implies that if a map, $S_{u,v}:\JJ\to \JJ $, is defined  by 
$S_{u,v}(y)=\lbrace u,v, y\rbrace$, then one finds,
\begin{equation}
\left[ S_{u,v}, S_{w,x}\right]=S_{w,\lbrace u,v,x\rbrace } -S_{\lbrace v,u,w\rbrace ,x }, 
\end{equation} 
so that the space of the linear maps, span$\lbrace S_{u,v}:u,v\in V\rbrace$,  is closed under a commutator bracket, and hence represents  a Lie algebra, $ \mathfrak{str(J)} $, termed to as  ``structure algebra''. Any  Jordan algebra induces a Jordan Triple System when we define the Jordan Triple product through
$
\lbrace u,v, w\rbrace= u \circ (v \circ w) - v \circ (u \circ w)
+ (u\circ v) \circ w 
 \ .
$ Moreover, introducing pairs, denoted by $\JJ$, and $\JJ^\ast $, of JTS, where  $\JJ$ and $\JJ^\ast $ are dual to each other, a linear map can be constructed 
amounting to, $\JJ \otimes \JJ^\ast\longrightarrow  \mathfrak{g} \mathfrak{l}(\JJ )  \oplus   
\mathfrak{g} \mathfrak{l}(\JJ^\ast)$, 
whose image is a Lie sub-algebra  
$\mathfrak{str (J)}$, and the Jordan identities  imply
the Jacobi identities for a graded Lie bracket on  $\JJ\oplus  \mathfrak{str (J)}\oplus \JJ^\ast$. Then, in  a graded algebra as the one in (\ref{3-gdrd}), the pair $\left( \mathfrak{g}_{+1}, \mathfrak{g}_{-1} \right) $ can be viewed as a Jordan pair, according to the correspondence,
$\lbrace x_\mp,y_\pm,z_\pm\rbrace _\pm =\left[ \left[ x_\mp,y_\pm \right], z_\pm \right]$.  
The procedure outlined above, termed to as  the  Kantor-Koecher-Tits correspondence,  when applied to the special Jordan algebra, $\JJ_2^C$, allows to interpret the $so(2,4)$ algebra in terms of Jordan triple systems and Jordan pairs and be cast as, 
$so(2,4)\simeq \mathfrak{co(J)}=\mathfrak g_{+1}\oplus \mathfrak g_0\oplus \mathfrak g_{-1}:=\JJ^\ast \oplus \mathfrak{str (J)}\oplus \JJ^\ast$,
meaning that the Abelian sub-algebra $\mathfrak g_{-1}(\mathfrak g_{+1})$ 
is generated in the space $\JJ(\JJ^\ast)$ (see \cite{Todor} for details about the conformal dynamical symmetry $so(2,4)$ of the hydrogen atom).  
The advantage of the Jordan algebra view on the conformal symmetry consists in the possibility of its straightforward  generalization to higher symmetry groups. Specifically in \cite{Dubois},\cite{Svetla}, attention has been drawn to the fact that the exceptional group $F_4$,   
hypothesized as internal space symmetry of a unified theory of strong and electroweak interactions, can be approached  over the octonion Jordan algebra $\JJ_3^O$.

\section{Summary}
We provided a concise review of recent progress on revealing the conformal-symmetry--color-neutrality connection in and near the infrared regime of  QCD  \cite{JAV_TP_MK_JHEP_2021}. The work  has been based on the assumption that  
the compactified boundary of the $AdS_5$ space, whose relevance to QCD follows from the $AdS_5/CFT_4$ duality conjecture, can be employed as  space  hosting the strong interaction degrees of freedom of QCD. As long as the topology of this space is  $S^1\times S^3$,  it contains a closed space-like hyper-surface ($S^3$ in this case), on which only charge neutral configurations of the type $2^n$ poles, all  necessarily neutral, can reside. In this way, the color-neutrality of hadrons finds a possible explanation. In addition, the space has the conformal group as isometry, a quality which allows to motivate the opening of the conformal window in the infrared by admitting the compactification radius as a second scale next to $\Lambda_{QCD}$. In consequence, the infrared regime acquires features of a perturbative one, permits for the approximation by Abelian color charges, and thus  allows one to conclude on the quantum mechanical limit of QCD as represented by the following  wave equation, 
\begin{eqnarray}
  \left[ \Box_{S^1\times S^3} -\alpha_sN_c\cot\chi +
    \frac{\alpha_s^2N_c^2}{4(K+1)^2} \right]\psi &=&0,
  \label{BiBo}
\end{eqnarray}
where $\Box_{S^1\times S^3}$ stands for the conformal wave operator on $S^1\times S^3$.
This wave equation describes quite realistically a broad range on hadron physics experiments from excitation spectra, over electromagnetic form factors, up to 
phase transitions. \\

\vspace{0.15cm}

\noindent
{\bf Acknowledgments}: This work has been partly supported by the Bulgarian National Science Fund, research grant DN 18/3.

\end {document}